\documentclass[11pt]{scrartcl}
\usepackage{graphicx}
\usepackage{bm}% bold math
\usepackage{hyperref}
\usepackage{color}

\usepackage{amsmath,amsfonts,amssymb}
\begin{document}

%%%%%%%%%%%%%%%%%%%%%%%%%%%%%%%%%%%%%%%%%%%

\def\o{\over}
\def\beq{\begin{align}}
\def\eeq{\end{align}}
\newcommand{\gsim}{ \mathop{}_{\textstyle \sim}^{\textstyle >} }
\newcommand{\lsim}{ \mathop{}_{\textstyle \sim}^{\textstyle <} }
\newcommand{\vev}[1]{ \left\langle {#1} \right\rangle }
\newcommand{\bra}[1]{ \langle {#1} | }
\newcommand{\ket}[1]{ | {#1} \rangle }
\newcommand{\EV}{ \text{eV} }
\newcommand{\KEV}{ \text{keV} }
\newcommand{\MEV}{ \text{MeV} }
\newcommand{\GEV}{ \text{GeV} }
\newcommand{\TEV}{ \text{TeV} }
\newcommand{\1}{\mbox{1}\hspace{-0.25em}\mbox{l}}
\newcommand{\headline}[1]{\noindent{\bf #1}}
\def\diag{\mathop\text{diag}\nolimits}
\def\Spin{\mathop\text{Spin}}
\def\SO{\mathop\text{SO}}
\def\O{\mathop\text{O}}
\def\SU{\mathop\text{SU}}
\def\U{\mathop\text{U}}
\def\Sp{\mathop\text{Sp}}
\def\SL{\mathop\text{SL}}
\def\tr{\mathop\text{tr}}
\def\mpl{M_\text{Pl}}

\def\dd{\mathrm{d}}
\def\ff{\mathrm{f}}
\def\BH{\text{BH}}
\def\inf{\text{inf}}
\def\ev{\text{evap}}
\def\eq{\text{eq}}
\def\SM{\text{sm}}
\def\Mpl{M_\text{Pl}}
\def\GeV{\text{GeV}}
\newcommand{\Red}[1]{\textcolor{red}{#1}}

%%%%%%%%%%%%%%%%%%%%%%%%%%%%%%%%%%%%%%%%%%%%%%%%%%%%%%%%%%%%%%%
\begin{titlepage}
\begin{center}

\hfill IPMU-15-0099 \\
\hfill \today

\vspace{1.5cm}
{\Large\bf 
Dynamics of Peccei-Quinn Breaking Field after Inflation\\
and\\
Axion Isocurvature Perturbations
}

\vspace{2.0cm}
{\bf Keisuke Harigaya}$^{(a,b)}$,
{\bf Masahiro Ibe}$^{(a, b)}$,\\
{\bf Masahiro Kawasaki}$^{(a, b)}$
and
{\bf Tsutomu T. Yanagida}$^{( b)}$

\vspace{1.0cm}
{\it
$^{(a)}${ICRR, University of Tokyo, Kashiwa, Chiba 277-8582, Japan}\\
$^{(b)}${Kavli IPMU (WPI), UTIAS, University of Tokyo, Kashiwa, Chiba 277-8583, Japan} 
}

\vspace{2.0cm}
\abstract{
The Peccei-Quinn mechanism suffers from the problem of the isocurvature perturbations.
The isocurvature perturbations are suppressed if the Peccei-Quinn breaking scale is large during inflation.
The oscillation of the Peccei-Quinn breaking field after inflation, however, leads to the formation of domain walls due to the parametric resonance effect.
In this paper, we discuss the evolution of the Peccei-Quinn breaking field after inflation in detail,
and propose a model where the parametric resonance is ineffective and hence domain walls are not formed.
We also discuss consistency of our model with supersymmetric theory.
}
\end{center}
\end{titlepage}
\setcounter{footnote}{0}

%\baselineskip 6mm

%%%%%%%%%%%%%%%%%%%
%---------------SECTION-------------------%

%%%%%%%%%%%%%%%%%%%%%%%%%%%%%%%%%%%
\section{Introduction}
%%%%%%%%%%%%%%%%%%%%%%%%%%%%%%%%%%%

The Peccei-Quinn (PQ) mechanism~\cite{Peccei:1977hh} is a compelling solution to the strong CP problem~\cite{'tHooft:1976up,Jackiw:1976pf,Callan:1976je}.
The mechanism not only solves the strong CP problem, but also predicts a Nambu-Goldstone boson associated with the spontaneous breaking of the PQ symmetry~\cite{Weinberg:1977ma,Wilczek:1977pj}, axion, which is a good candidate of dark matter~\cite{Preskill:1982cy,Abbott:1982af,Dine:1982ah}.

The PQ mechanism, however, brings about cosmological problems.
If the PQ symmetry is restored during inflation, the breaking of the PQ symmetry after inflation leads to formation of domain walls~\cite{Sikivie:1982qv}.
Unless the domain wall number is unity, the domain walls are stable and hence eventually  dominate the energy density of the universe.
If the PQ symmetry is broken during inflation, on the other hand, the quantum fluctuations of the axion induce the isocurvature perturbations of cold dark matter~\cite{Linde:1984ti,Seckel:1985tj,Lyth:1989pb,Turner:1990uz,Linde:1991km}, which is strictly constrained by observations of the cosmic microwave background (CMB)~\cite{Ade:2015lrj}.
This is called the isocurvature perturbation problem.
To evade the constraint,
%from observations of the CMB,
the Hubble scale during inflation must be smaller than about $10^{7}$ GeV (see Sec.~\ref{sec:review}).
This strict bound is incompatible with many inflation models including chaotic inflation~\cite{Linde:1983gd}, which is the simplest inflation model without the initial condition problem and predicts a large Hubble scale during inflation, $H_\text{inf} \sim \text{several}\times 10^{13}$ GeV.

As was first pointed in~\cite{Linde:1991km} (see also~\cite{Choi:2014uaa,Chun:2014xva,Fairbairn:2014zta,Nakayama:2015pba}), the isocurvature perturbations are suppressed if the PQ breaking scale is large during inflation.
Due to the large PQ breaking scale, the fluctuations of the misalignment angle are  suppressed and hence the axion density perturbations are also suppressed.%
\footnote{
See Refs.~\cite{Higaki:2014ooa,Kitajima:2014xla,Kawasaki:2015lea} for other possible solutions to the isocurvature perturbation problem.
}
For this suppression mechanism to work, the large expectation value of the PQ breaking field should decrease to the present value after inflation. 
However, the relaxation of the PQ breaking scale may lead to the restoration of the PQ symmetry in the following way.
After inflation, the PQ breaking field oscillates around the origin, which produces large fluctuations of the PQ breaking field through the parametric resonance effect~\cite{Kofman:1994rk}.
The large fluctuations lead to the restoration of the PQ symmetry and hence the formation of domain walls~\cite{Tkachev:1995md,Kasuya:1996ns,Kasuya:1997ha,Tkachev:1998dc,Kawasaki:2013iha}.
To prevent the formation of domain walls, the PQ breaking scale during inflation cannot be arbitrary large.
The recent study with lattice simulations~\cite{Kawasaki:2013iha} shows that  the upper bound on the Hubble scale from the isocurvature perturbations is raised to at most $10^{12}$ GeV
when the axion is the dominant component of dark matter.

In this paper, we discuss the evolution of the PQ breaking field after inflation in detail, and propose a model where the parametric resonance is ineffective and hence domain walls are not formed.
Thus, the model allows for both axion dark matter and inflation models with large Hubble scale.
We also discuss consistency of our model with supersymmetric theory.

This paper is organized as follows.
In Sec.~\ref{sec:review}, we review the isocurvature perturbation problem of the PQ mechanism and suppression of isocurvature perturbations by a large PQ breaking scale during inflation.
In Sec.~\ref{sec:relaxation}, we discuss the evolution of the PQ breaking field and restoration of PQ symmetry after inflation.
We propose a model of the PQ symmetry breaking without the restoration of the PQ symmetry.
The final section is devoted to summary and discussion.

%%%%%%%%%%%%%%%%%%%%%%%%%%%%%%%%%%%%%%%%%%%%%%%%%%%%%%%%%%%%%%%%%%%%%%%%%%
\section{Isocurvature perturbation problem and a large PQ breaking scale}
\label{sec:review}
%%%%%%%%%%%%%%%%%%%%%%%%%%%%%%%%%%%%%%%%%%%%%%%%%%%%%%%%%%%%%%%%%%%%%%%%%%

In this section,
we review the problem of isocurvature perturbations in the PQ mechanism and suppression of their amplitude by a large scale of the PQ breaking during inflation.

%%%%%%%%%%%%%%%%%%%%%%%%%%%%%%%%%%%%%%%%%%%%%%%%%%%%%%%%%%%%%%%%%%%

\subsection{QCD axion as a dark matter candidate}

The axion $a$ is the Nambu-Goldstone boson associated with the spontaneous breaking of the PQ symmetry~\cite{Peccei:1977hh,Weinberg:1977ma,Wilczek:1977pj}.
It is assumed that the PQ symmetry has anomaly of QCD. Then the axion obtains its potential by non-perturbative QCD dynamics,
\begin{equation}
   V(a) \sim m_{\pi}^2 f_{\pi}^2 (1 - \text{cos}\theta_a),
   ~~~~~\theta_a \equiv a/f_a, 
\end{equation}
where $m_\pi$, $f_\phi$ and $f_a$ are the mass of the pion, the pion decay constant, and the axion decay constant.
Here, we have shifted the axion so that its vacuum expectation value (VEV) is zero. 

In this paper, we assume that the PQ symmetry is spontaneously broken during inflation
to avoid the domain wall problem.
%so that
%the axion takes a uniform field value within the observable universe.
%domain walls are inflated away.
Then in the early universe, the axion in general has a non-zero field value $a_i$.
After inflation, around a temperature of the QCD scale, the axion begins to oscillate and behaves as cold dark matter.
The present density of the axion is given by~\cite{Bae:2008ue}
\begin{align}
\Omega_a h^2 = 0.2 \times \theta_i^2\left(\frac{f_a}{10^{12}~\text{GeV}}\right)^{1.19},
\end{align}
where $\theta_i \equiv a_i/ f_a$ is the initial misalignment angle.
For $f_a = 10^{11\mathchar`-12}$ GeV, the density of the axion oscillation is as large as that of dark matter.

%%%%%%%%%%%%%%%%%%%%%%%%%%%%%%%%%%%%%%%%%%%%%%%%%%%%%%%%%%%%%%%%%%%

\subsection{Isocurvature perturbation}

The initial axion field value $a_i$ takes almost the same value in the whole observable universe.
However, quantum effect during inflation inevitably induces fluctuations of the initial field value, $\delta a_i \sim H/ (2\pi)$~\cite{Mukhanov:1981xt,Hawking:1982cz,Starobinsky:1982ee,Guth:1982ec,Bardeen:1983qw}.
Taking into account the fluctuations, the density parameter of the axion is modified as
\begin{align}
    \Omega_a h^2 &= 0.2 \times \theta_{i,\text{eff}}^2
    \left(\frac{f_a}{10^{12}~\text{GeV}}\right)^{1.19}, \nonumber \\
    \theta_{i,\text{eff}}^2 & \equiv  \theta_i^2 +\delta \theta_i^2=
    \theta_i^2 + \left( \frac{H_\text{inf}}{2\pi f_a} \right)^2.
\end{align}
The axion fluctuations are independent of those of the inflaton field in the flat time slice.
Thus, the fluctuations of the axion result in un-correlated isocurvature perturbations of cold dark matter~\cite{Linde:1984ti,Seckel:1985tj,Lyth:1989pb,Turner:1990uz,Linde:1991km}.
The power spectrum of the isocurvature perturbations is given by
\begin{equation}
   \mathcal{P}_{S_c} = \left(\frac{\delta \Omega_a}{ \Omega_c}\right)^2 = 
   \frac{4}{\theta_{i,\text{eff}}^2} \left(\delta \theta_{i} \right)^2 
   \left(\frac{\Omega_a}{\Omega_c}\right)^2=
   \frac{4}{\theta_\text{i,{eff}}^2} 
   \left( \frac{H_\text{inf}}{2\pi f_a} \right)^2 
   \left(\frac{\Omega_a}{\Omega_c}\right)^2,
\end{equation}
where $\Omega_c$ is the density parameter of cold dark matter.

\begin{figure}[htbp]
\centering
\includegraphics[width=0.6\linewidth]{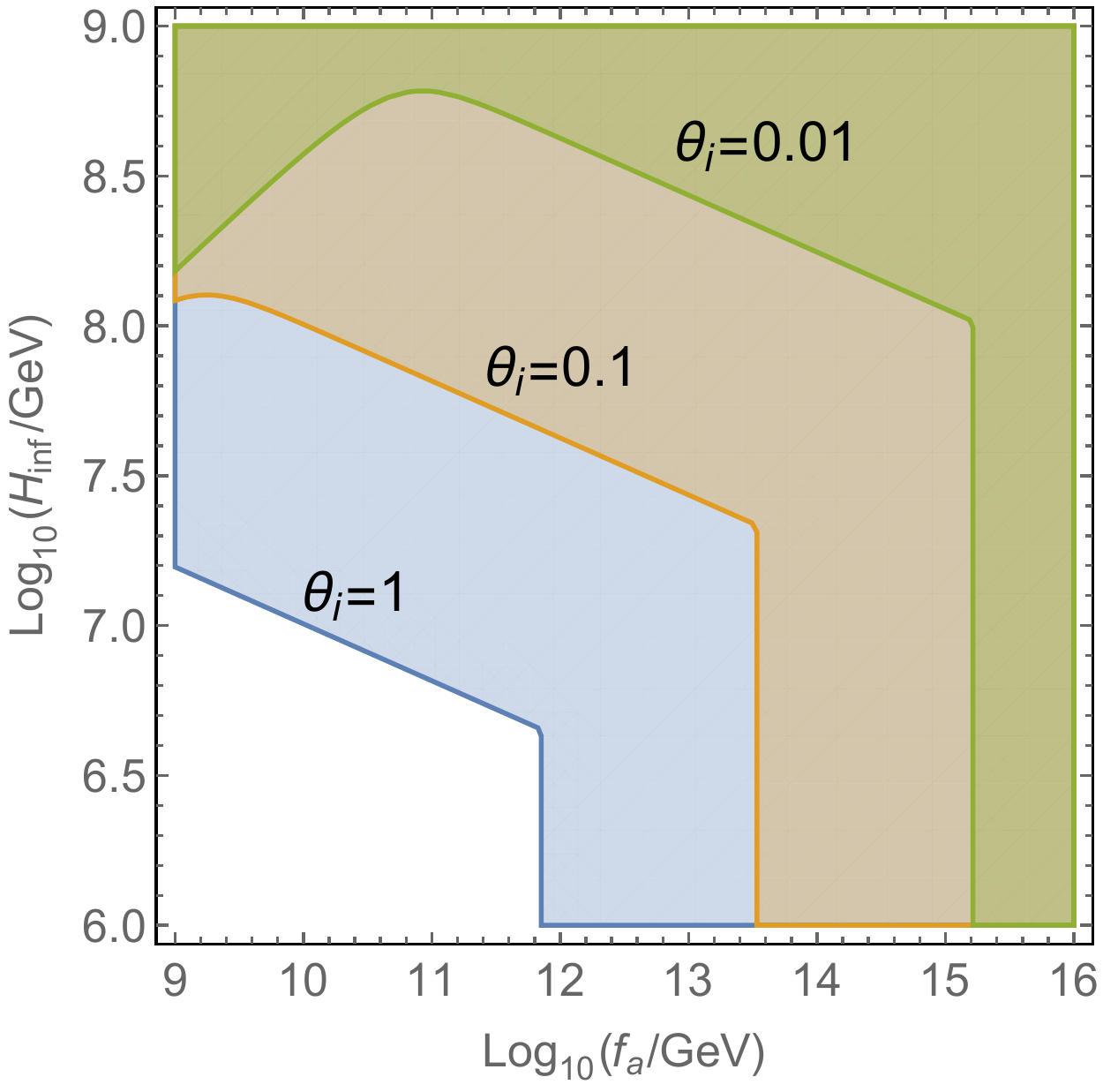}
\caption{\sl \small
Bound on the axion decay constant $f_a$ and the Hubble scale during inflation $H_\text{inf}$ for $\theta_i = 1$, $0.1$ and $0.01$.
Shaded regions are excluded by the constraint on the isocurvature perturbations, $\alpha_c < 0.033$, or by too large cosmic density of the axion.
}
\label{fig:fa_Hinf}
\end{figure}

Uncorrelated isocurvature perturbations of cold dark matter are constrained as
$\alpha_c \equiv \mathcal{P}_{S_c}/ (\mathcal{P}_\zeta+\mathcal{P}_{S_c}) < 0.033$, where $\mathcal{P}_\zeta \simeq 2\times 10^{-9}$ is the power spectrum of the curvature perturbations~\cite{Ade:2015lrj}.
In Figure~\ref{fig:fa_Hinf},
we show the bound on the axion decay constant $f_a$ and the Hubble scale during inflation $H_\text{inf}$ for $\theta_i = 1$, $0.1$ and $0.01$.
Shaded regions are excluded by the constraint from the isocurvature perturbations or too large cosmic density of the axion.
We do not show the parameter region with $f_a < 10^9$~GeV because astrophysical constraints exlude that region.
The Hubble scale during inflation must be smaller than about $10^7$ GeV for $\theta_i = O(1)$,
which severely constrains inflation models.
For example, the simplest inflation model without the initial condition problem, namely chaotic inflation~\cite{Linde:1983gd},
predicts a large Hubble scale $H_\text{inf} \sim \text{several}\times 10^{13}$ GeV.

%%%%%%%%%%%%%%%%%%%%%%%%%%%% Figure 1 %%%%%%%%%%%%%%%%%%%%%%%%%%%%%%%%%
%%%%%%%%%%%%%%%%%%%%%%%%%%%%%%%%%%%%%%%%%%%%%%%%%%%%%%%%%%%%%%%%%%%%%%%%

%%%%%%%%%%%%%%%%%%%%%%%%%%%%%%%%%%%%%%%%%%%%%%%%%%%%%%%%%%%%%%%%%%%

\subsection{Large PQ breaking field value during inflation}

The isocurvature perturbations can be suppressed if the PQ breaking scale is larger than the present one during inflation~\cite{Linde:1991km}.
Suppose that the PQ breaking field $P$ takes a large field value $P_\text{inf}$ during inflation, which we take to be real by a PQ rotation.
The large field value is explained by a negative Hubble induced mass of the PQ breaking field during inflation.
Then the decay constant of the axion during inflation $f_\text{inf}$ is given by
\begin{align}
f_\text{inf} = \sqrt{2} P_\text{inf} / N_\text{DW},
\end{align}
where $N_\text{DW}$ is the domain wall number of the axion model.
Then the fluctuation of the initial misalignment angle is written as
\begin{align}
\delta \theta_i = \frac{H_\text{inf}}{2\pi f_\text{inf}} = \frac{H_\text{inf}}{2\pi \sqrt{2}P_\text{inf}}\times N_\text{DW},
\end{align}
which leads to the power spectrum of the isocurvature perturbations as
\begin{align}
{\cal P}_{S_c} =
\frac{4}{\theta_{i,\text{eff}}^2} \left( \frac{ N_\text{DW}H_\text{inf }}{2\pi \sqrt{2}P_\text{inf}} \right)^2  \left(\frac{\Omega_a}{\Omega_c}\right)^2.
\end{align}

\begin{figure}[htbp]
\centering
\includegraphics[width=0.6\linewidth]{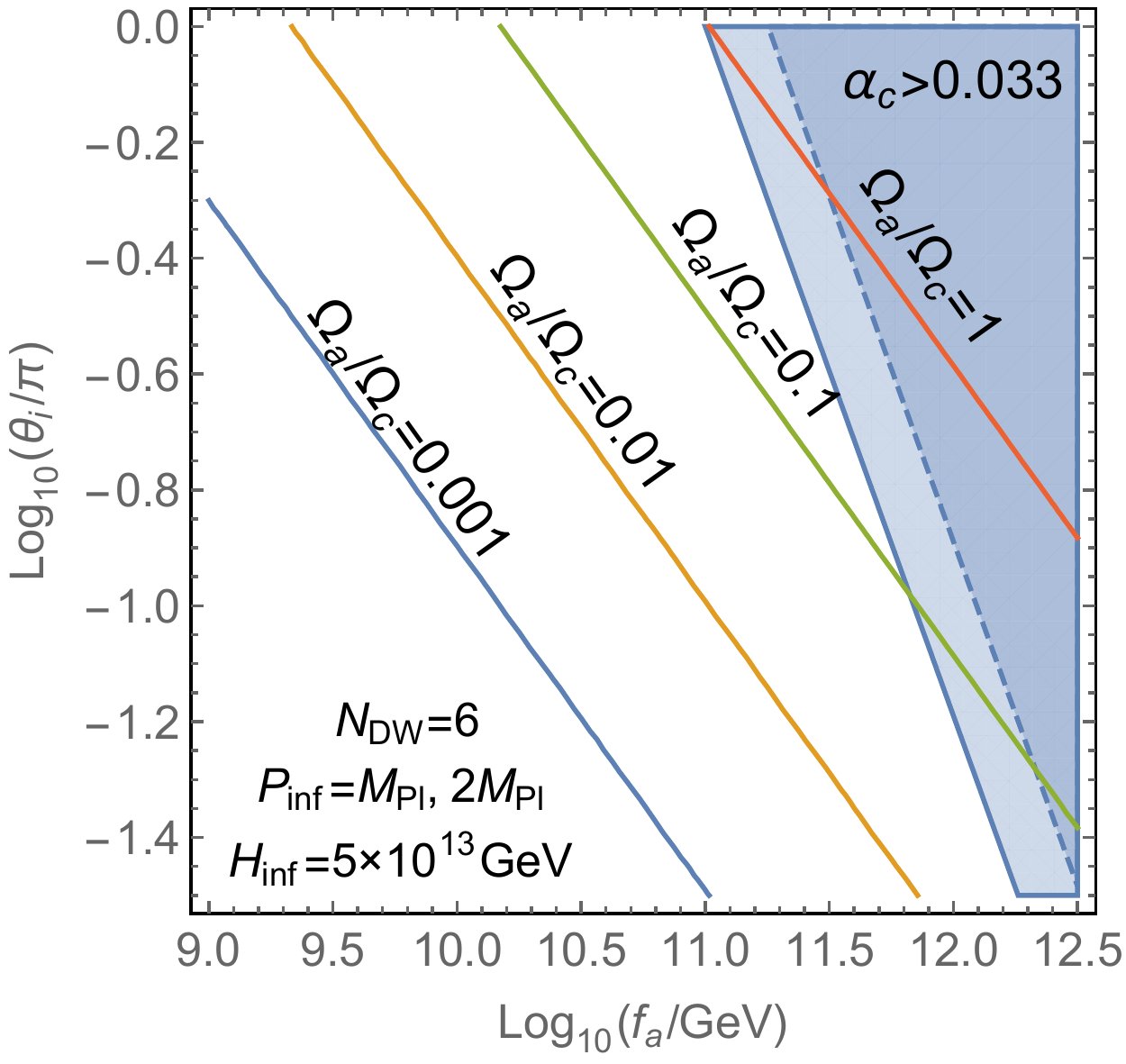}
\caption{\sl \small
Bound on the axion decay constant $f_a$ and the initial misalignment angle $\theta_i$ for
$N_\text{DW}=6$,
$H_\text{inf} = 5\times 10^{13}$ GeV.
The shaded region with a solid (dashed) line boundary shows the constraint from the isocurvature perturbation for $P_\text{inf} =(2) \mpl$.
}
\label{fig:fa_theta}
\end{figure}

In Figure~\ref{fig:fa_theta}, we show the bound on the axion decay constant $f_a$ and the initial misalignment angle $\theta_i$ for
$N_\text{DW}=6$ (DFSZ model~\cite{Dine:1981rt,Zhitnitsky:1980tq}),
$H_\text{inf} = 5\times 10^{13}$ GeV (tensor-to-scalar ratio $r\simeq 0.04$) and $P_\text{inf} = \mpl$ as well as $2\mpl$.
Even if the Hubble scale during inflation is such large,
for $f_{a} \sim 10^{11}$ GeV and $\theta_i = O(1)$, the axion can be a sizable component of dark matter while the bound from the isocurvature perturbations is avoided.
It should be remarked that the Hubble scale during inflation can be determined by measuring the tensor-to-scalar ratio $r$, whose present bound is $r \lesssim 0.1$~\cite{Ade:2015lrj}.
Future ground based experiments can detect  $r\gtrsim 0.003$~\cite{Creminelli:2015oda}, which corresponds to $H_{\rm inf}\gtrsim 10^{13}$~GeV.
Therefore, it is important
to examine whether an axion model is
consistent with high scale inflation with $H_\text{inf}\gtrsim 10^{13}$~GeV.

%%%%%%%%%%%%%%%%% Figure 2 %%%%%%%%%%%%%%%%%%%%%%%%%%
%%%%%%%%%%%%%%%%%%%%%%%%%%%%%%%%%%%%%%%%%%%%%%%%%%%%%%%%%%

%%%%%%%%%%%%%%%%%%%%%%%%%%%%%%%%%%%%%%%%%%%%%%%%%%%%%%%%%%%%%%%%%%%
\section{Evolution of PQ breaking field}
\label{sec:relaxation}
%%%%%%%%%%%%%%%%%%%%%%%%%%%%%%%%%%%%%%%%%%%%%%%%%%%%%%%%%%%%%%%%%%%

In the previous section, we have shown that the bound from the isocurvature perturbations is evaded if the field value of the PQ breaking field is large during inflation.
In this section, we discuss the evolution of the PQ breaking field after inflation in detail.
We show that the oscillation of the PQ breaking field driven by a quartic potential leads to the formation of domain walls.
Then we propose a model where domain walls are not formed.
We also discuss the embedding of the model to supersymmetric theory.

%%%%%%%%%%%%%%%%%%%%%%%%%%%%%%%%%%%%%%%%%%%%%%%%%%%%%%%%%%%%%%%%%%%

\subsection{General discussion}
\label{sec:analytic}

Let us discuss the dynamics of the PQ breaking field analytically, following Ref.~\cite{Dine:1995kz}.
We assume the following potential of the PQ breaking field $P$ and the inflaton $\phi$,
\begin{align}
\label{eq:potential}
%V(\chi) = \frac{\lambda_{2n}^2}{2^n} \chi^{2n} - \frac{c_H}{2} H^2 \chi^2,
V(\phi,P) = V_\text{inf}(\phi) +  \lambda_{2n}^2 |P|^{2n} - \frac{c_H}{3} V_\text{inf}(\phi) |P|^2,
\end{align}
where $V_\text{inf}(\phi)$ is the potential of the inflaton, and
%$\chi$ is the radial direction of the PQ breaking field, $P = \chi/ \sqrt{2}$,
$\lambda_{2n}$ as well as $c_H (>0)$ are constants.%
%, and $H$ is the Hubble scale.%
\footnote{
Since the inflaton potential $V_\text{inf}(\phi)$ is neutral under any symmetries, it is natural that the interaction between the inflaton and the PQ breaking field in Eq.~(\ref{eq:potential}) is present. This is also the case even in supersymmetry theory.}
Here, we neglect the mass term of the PQ breaking field, which is important only when the field value of the PQ breaking field becomes sufficiently small.
%The Hubble induced mass term is given by a coupling between the PQ breaking field and the inflaton,%
%\begin{align}
%\label{eq:Hubble induced mass}
%{\cal L}_\text{int} = + \frac{c_H}{3} V(\phi)_\text{inf} |P|^2.
%\end{align}
With this potential, the PQ breaking field obtains a large field value during inflation,
\begin{align}
\label{eq:PQ inf}
P_\text{inf} = \left( \frac{c_H H_\text{inf}^2}{n\lambda_{2n}^2}  \right)^{\frac{1}{2(n-1)}},
\end{align}
where $H_\text{inf}$ is the Hubble scale during inflation.

After inflation, the inflaton begins oscillation around the origin.
During the oscillatory phase, the PQ breaking field may also begin oscillation around the origin.
The equation of motion of the radial direction of $P$, $ \chi = |P|/\sqrt{2}$, is
\begin{align}
\ddot{\chi} + 3 H \dot{\chi} + \frac{2n \lambda_{2n}^2}{2^n} \chi^{2n-1} - \frac{c_H}{2}H^2 \chi =0,
\end{align}
%%
%Here, we consider the evolution of the zero mode.
where $H$ is the Hubble scale.
Here, we replace $V_\text{inf}(\phi)$ with its time averaged value, $3H^2 /2 $, which is a good approximation as long as the mass of the inflaton is much larger than the Hubble scale.

It is convenient to use the number of e-folding $N\equiv \text{ln}(R/ R_i)$ as a time variable, where $R$ is the scale factor and $R_i$ is its initial value when the inflaton starts oscillation.
In the oscillatory phase of the inflaton, namely in a matter dominant universe,
the equation of motion of $\chi$ is then given by
\begin{align}
\frac{\text{d}^2}{\text{d}N^2} \chi + \frac{3}{2}\frac{\text{d}}{\text{d}N} \chi + \frac{2n \lambda_{2n}^2}{2^nH^2} \chi^{2n-1} - \frac{c_H}{2}\chi =0.
\end{align}
We further rewrite the equation of motion by $\psi \equiv \chi \text{exp}(\frac{3N}{2(n-1)})$,
\begin{align}
\label{eq:eom reduced}
\frac{\text{d}^2}{\text{d}N^2} \psi + \frac{3(n-3)}{2(n-1)}\frac{\text{d}}{\text{d}N} \psi - \left(\frac{9(n-2)}{4(n-1)^2} +\frac{c_H}{2} \right)\psi + \frac{2n\lambda_{2n}^2}{2^nH_i^2} \psi^{2n-1}=0,
\end{align}
where $H_i$ is the Hubble scale when the inflaton starts oscillation.
Finally, we normalize the field $\psi$ so that the initial value of the normalized field is of $O(1)$,
\begin{align}
\label{eq:eom norm}
\frac{\text{d}^2}{\text{d}N^2} \rho + \frac{3(n-3)}{2(n-1)}\frac{\text{d}}{\text{d}N} \rho
- \left(\frac{9(n-2)}{4(n-1)^2} +\frac{c_H}{2} \right)\rho + \frac{c_H}{2} \rho^{2n-1}=0,\nonumber \\
\psi \equiv \left( \frac{2^nc_HH_i^2}{4n\lambda_{2n}^2} \right)^{\frac{1}{2n-2}} \rho.
\end{align}
Note that all coefficients in Eq.~(\ref{eq:eom norm}) is $N$-independent.

The equation of motion of $\rho$ is nothing but that of a particle with a potential
\begin{align}
V(\rho) = - \frac{1}{2}\left(\frac{9(n-2)}{4(n-1)^2} +\frac{c_H}{2} \right)\rho^2 + \frac{c_H}{4n} \rho^{2n},
\end{align}
whose minimum is at
\begin{align}
\rho = \left(   1 + \frac{9(n-2)}{2(n-1)^2 c_H}  \right)^{\frac{1}{2n-2}} \equiv \rho_0.
\end{align}
The initial value of $\rho$, $\rho_i$, when the inflaton starts oscillation is expected be of $O(1)$.
Its definite value, however, depends on
the dynamics of the inflaton between the inflationary phase and the oscillatory phase.

For $n=2$, the friction term has a wrong sign, and hence the field value of $\rho$ grows and eventually starts oscillation around the origin. 
We will cross-check this conclusion numerically later.
Due to the oscillation, the parametric resonance effect grows the fluctuations of the PQ breaking field exponentially in time~\cite{Kofman:1994rk}.
The PQ field with large initial field value oscillates for a long time until  it settles down to the minimum of the potential.
Thus, the oscillation produces  large fluctuations, which eventually leads to the restoration of the PQ symmetry and hence the formation of domain walls~\cite{Tkachev:1995md,Kasuya:1996ns,Kasuya:1997ha,Tkachev:1998dc,Kawasaki:2013iha}.

\begin{figure}[htbp]
\centering
\includegraphics[width=0.6\linewidth]{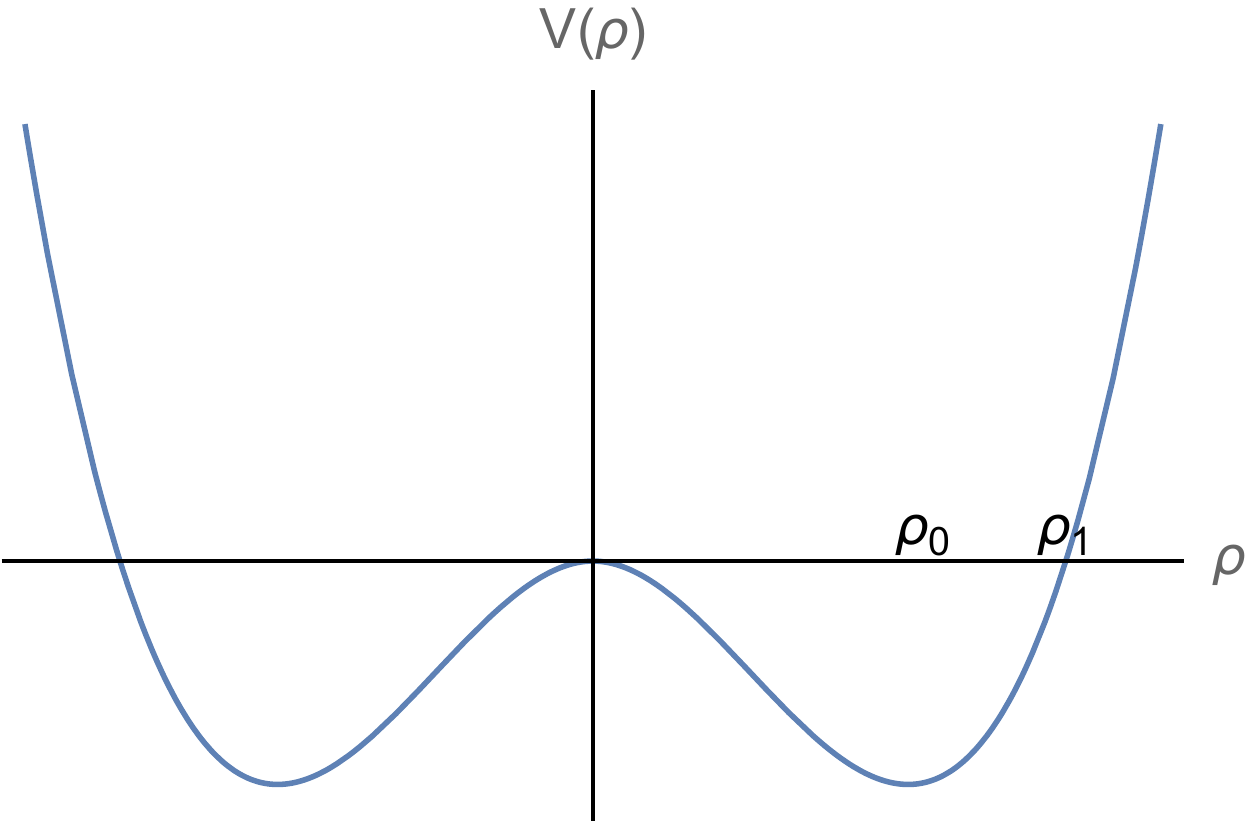}
\caption{\sl \small
The potential of $\rho$.}
\label{fig:rho potential}
\end{figure}

For $n=3$, the friction term is absent and hence $\rho$ oscillates continuously without attenuation.
The potential of $\rho$ is depicted in Figure~\ref{fig:rho potential}.
If $\rho_i$ is within the range $|\rho_i|< \rho_1 \equiv [(27/8) c_H^{-1} +3]^{1/4}$,
$\rho$ oscillates around $\rho_0$.
If $|\rho_i| > \rho_1$, $\rho$ oscillates around the origin.
In both cases, the PQ breaking field oscillation might produce fluctuations.%
\footnote{
See also Ref.~\cite{Chun:2014xva} for discussion on the parametric resonance in a sextet potential,
although the oscillation of the PQ breaking field is not taken into account.
}
So let us consider the evolution of the fluctuations of the PQ breaking field around the zero mode for $n=3$.
The evolution is described by
\begin{align}
   \frac{\text{d}^2}{\text{d}N^2} \delta \rho_k 
   + \frac{k^2}{H_i^2} e^N \delta \rho_k
   - \left(\frac{9}{16} +\frac{c_H}{2} \right)\delta \rho_k 
   + \frac{5}{2} c_H \rho^{4}\delta \rho_k=0,
   \label{eq:evol_fluc}
\end{align}
where $k$ is a comoving wave number of the fluctuation $\delta \rho_k$.
The strength of the parametric resonance is expected to be estimated by the magnitude of the coefficient $5c_H \rho^4 /2$,
in analogy with the $q$ parameter of the Mathieu's equation.
In the present case, the magnitude is of $O(1)$.
It is  expected that resonance bands are not broad.
Since effective wave numbers evolve in proportional to $\text{exp}(N/2)$~[see, the 2nd term in Eq.~(\ref{eq:evol_fluc})], given modes are in resonance bands at most during one oscillation. Thus the parametric resonance is considered to be ineffective.
We have confirmed this by numerically solving the evolution equation. 

% We have calculated the evolution of the zero mode $\rho$ and the perturbation $\delta \rho_k$ numerically,
% and show the evolution of the perturbation in Figure~\ref{fig:perturbation}.
% Initially, the perturbation glows only linearly in $N$.
% After the wave number term becomes important, the fluctuation dumps.
% As expected, the parametric resonance is absent.

For $n>3$, the friction term has a correct sign.
Thus,  $\rho$ quickly relaxes to the minimum $\rho_0$.
The PQ symmetry is not restored.

%%%%%%%%%%%%%%%%%%%%% Figure 3 %%%%%%%%%%%%%%%%%%%%%%%
%%%%%%%%%%%%%%%%%%%%%%%%%%%%%%%%%%%%%%%%%%%%%%%%%%%%%%%

% \begin{figure}[htbp]
% \centering
% \includegraphics[width=0.7\linewidth]{perturbation.pdf}
% \caption{\sl \small
% The evolution of the perturbation of $\rho$.
% We normalize the perturbation by its initial value, $\delta \rho_k (N = 0)$.
% }
% \label{fig:perturbation}
% \end{figure}

%%%%%%%%%%%%%%%%%%%%%%%%%%%%%%%%%%%%%%%%%%%%%%%%%%%%%%%%%%%%%%%%%%%

\subsection{More on Quartic potential}

Let us discuss the case with $n=2$, i.e.~a quartic potential in detail,
including the evolution of the inflaton from the inflationary phase to the oscillatory phase.
To be definite, we assume the inflaton potential around the origin,
\begin{align}
V_\text{inf}(\phi) = \frac{1}{2}m^2 \phi^2,
\end{align}
where $\phi$ is the inflaton and $m$ is its mass.
Then the equations of motion of the inflaton and the PQ breaking field are given by
\begin{align}
\ddot{\phi} + 3 H \dot{\phi} + m^2 \phi =0,\\ 
%-\frac{c_H}{3} m^2 \phi \chi^2 =0,\\
\ddot{\chi} + 3 H \dot{\chi} + \lambda_4^2 \chi^3 -\frac{c_H}{3} m^2 \phi^2 \chi =0.
\end{align}
%%
%where $\chi = \sqrt{2}P $.
Here, we neglect the back reaction of the evolution of the PQ breaking field on the dynamics of the inflaton.

We rewrite the equation of motion of $\chi$ to a more convenient form.
For an initial condition of $\phi = \phi_0$, the field value of $\chi$ determined by Eq.~(\ref{eq:PQ inf}) is
\begin{align}
\chi_0 = \sqrt{\frac{c_H}{6}}\frac{m\phi_0}{\lambda_4}.
\end{align}
Then the equation of motion of $r\equiv \chi / \chi_0$ is
\begin{align}
\ddot{r} + 3 H \dot{r} +\frac{1}{6} c_H m^2 \phi_0^2 r^3 -\frac{c_H}{3} m^2 \phi^2 r =0,
\end{align}
with the initial condition $r_0=1$.
One can see that the dynamics of $r = \chi / \chi_0$ does not depend on $\lambda_4$.

\begin{figure}[htbp]
\centering
\includegraphics[width=0.7\linewidth]{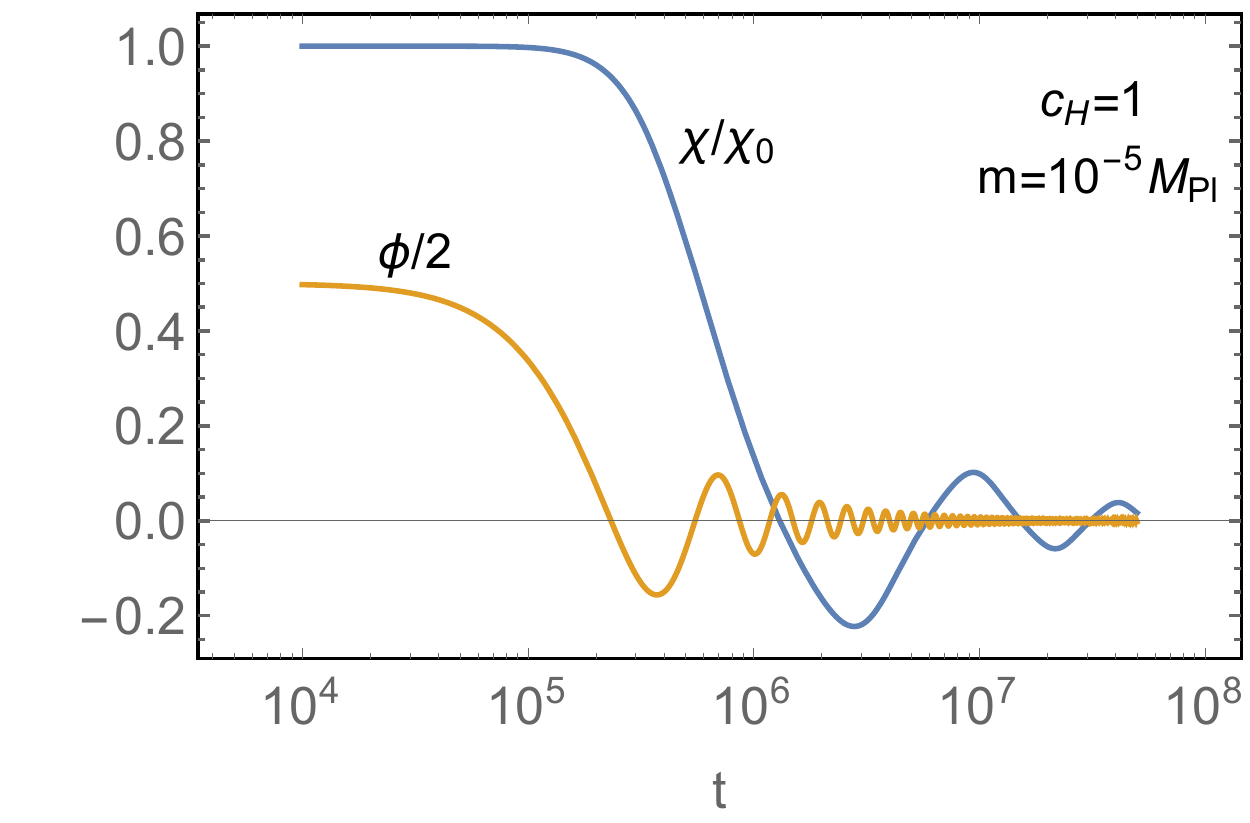}
\caption{\sl \small
The evolution of the PQ breaking field after inflation for the quartic potential.
We also show the evolution of the inflaton.
}
\label{fig:osc_4}
\end{figure}

In Figure~\ref{fig:osc_4}, we show the evolution of the PQ breaking field for
$m=10^{-5}\mpl$, $c_H=1$ with the unit system where the reduced Planck scale is unity.
Here, we take the initial condition $\phi_0 = \mpl$.
It can be seen that the PQ breaking field starts oscillation around the origin, which is consistent with our analytical estimation.

%%%%%%%%%%%%%%%%%%%%%% Figure 4 %%%%%%%%%%%%%%%%%%%%%%%%%%%
%%%%%%%%%%%%%%%%%%%%%%%%%%%%%%%%%%%%%%%%%%%%%%%%%%%%%%%%%%%%%

As the PQ breaking field oscillates with the quartic potential, the oscillation produces the fluctuations of the PQ breaking field by the parametric resonance effect~\cite{Kofman:1994rk}.
If the oscillation lasts long, the fluctuations eventually restore the PQ symmetry and hence domain walls are formed~\cite{Tkachev:1995md,Kasuya:1996ns,Kasuya:1997ha,Tkachev:1998dc,Kawasaki:2013iha}.
In order to prevent the formation of domain walls, the field value of the PQ breaking field when it start oscillation, $P_\text{osc}$, must satisfy~\cite{Kawasaki:2013iha}
\begin{align}
   P_\text{osc} < 10^{4} \vev{P}
   = 10^{16}~\text{GeV} \times \frac{\vev{P}}{10^{12}\text{GeV}},
\end{align}
where $\vev{P} (=N_\text{DW}f_a/\sqrt{2})$ is the present vacuum expectation value.

\begin{figure}[htbp]
\centering
\includegraphics[width=0.45\linewidth]{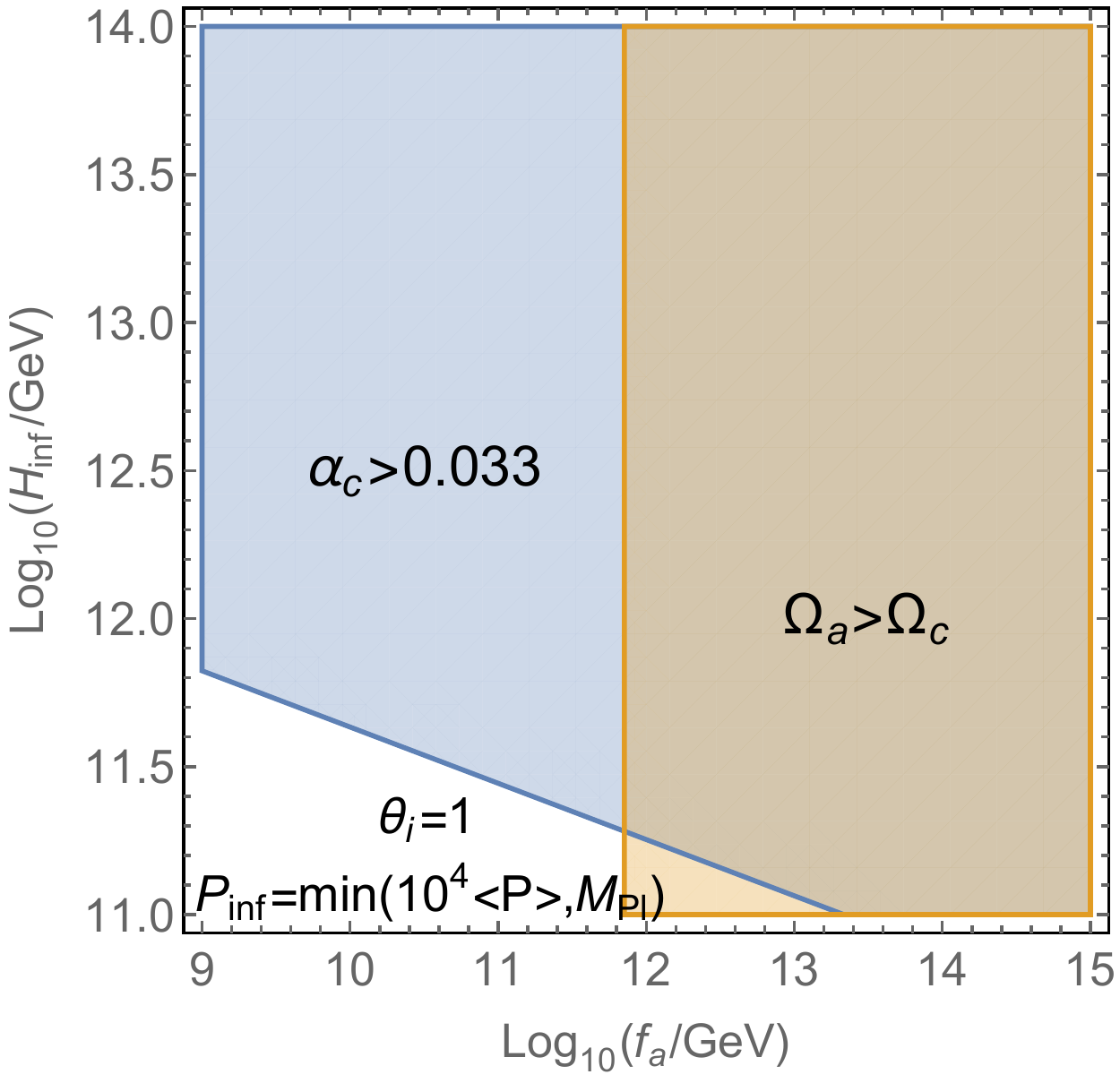}\hspace{5mm}
\includegraphics[width=0.45\linewidth]{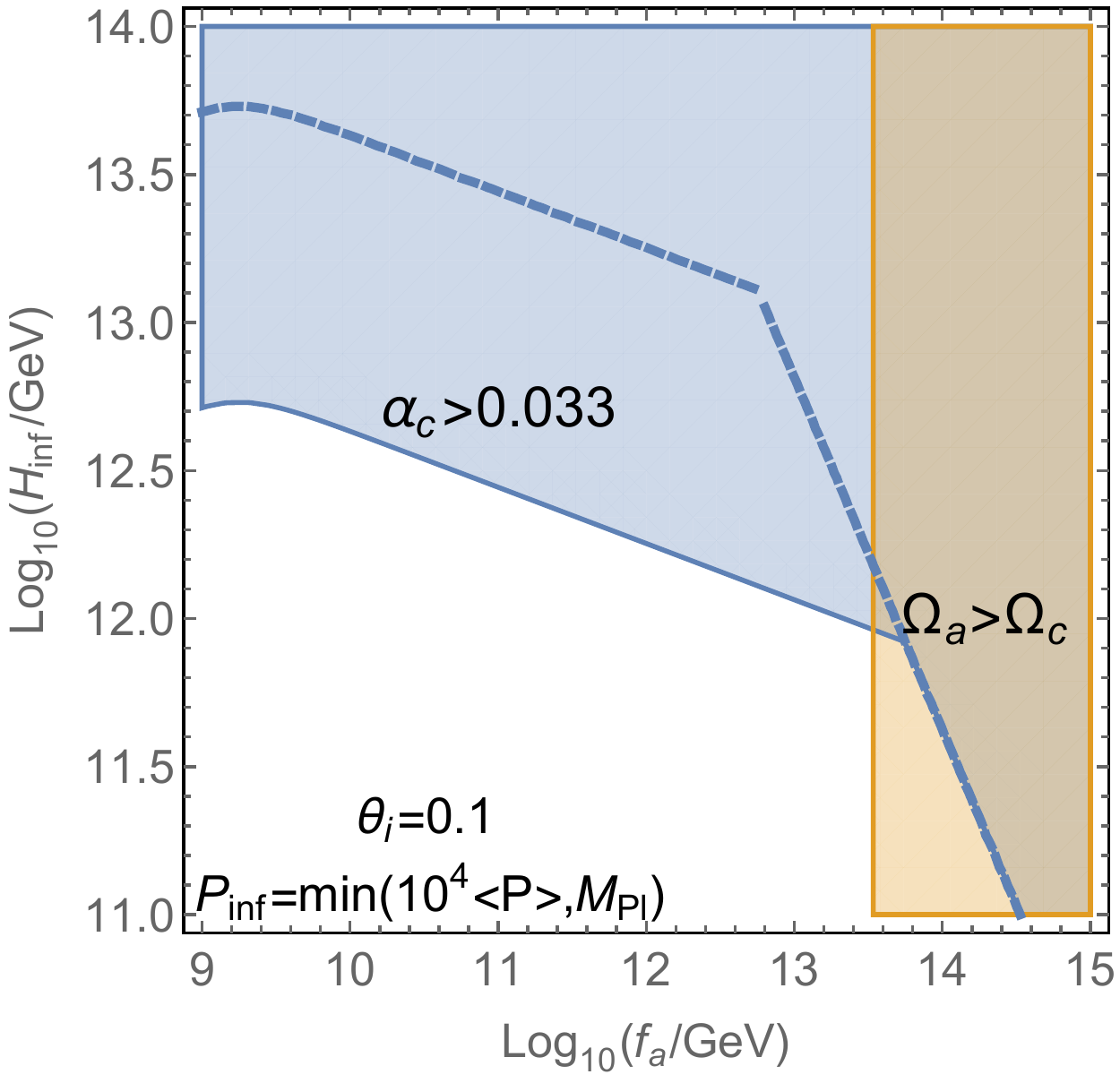}
\caption{\sl \small
The constraint on the axion decay constant $f_a$ and the Hubble scale during inflation $H_\text{inf}$ for
the quartic potential.
Here, we take $P_\text{inf}=$min$(10^{4} \vev{P},\mpl)$.
The dashed line in the right panel shows the upper bound on $H_\text{inf}$ for
$P_\text{inf}=$min$(10\times 10^{4} \vev{P},\mpl)$.
}
\label{fig:fa_Hinf_4}
\end{figure}

In Figure~\ref{fig:fa_Hinf_4}, we show the constraint on 
the axion decay constant $f_a$ and the Hubble scale during inflation $H_\text{inf}$
when the field value of the PQ breaking field during inflation $P_\text{inf}$ is as large as min$(10^{4} \vev{P},\mpl)$.
Here, we require that $P_\text{inf} < \mpl$ so that the PQ breaking field does not dominate the potential energy during inflation.
For $\theta_i = 1$, the Hubble scale during inflation must be smaller than $10^{12}$ GeV.
The constraint is more stringent $(H_\text{inf} < 2\times 10^{11}~\text{GeV})$ if we require that axion is dark matter. 
For $\theta_i = 0.1$, a larger Hubble scale is allowed due to smaller abundance of the axion.
For $\theta_i = 0.1$, we also show the upper bound on the Hubble scale with $P_\text{inf}=$ min$(10\times 10^{4} \vev{P},\mpl)$ by a dashed line.
We multiply the factor of $10$ to take into account a larger Hubble scale when the CMB scale exits the horizon than that at the end of the inflation by a factor of about $10$ in typical large field inflation models.
(Thus, this loose upper bound should not be applied to small field models.)
In this case the Hubble scale close to $10^{14}$ GeV is allowed although $H_\text{inf} < 10^{12}$~GeV is necessary for axion to account for dark matter.  
Therefore, even if we take small $\theta_i$ and adopt the loose bound, the Hubble scale during inflation should be less than $10^{12}$~GeV when the axion is a dominant component of dark matter.

%%%%%%%%%%%%%%%%%%%%%% Figure 5  %%%%%%%%%%%%%%%%%%%%%%%%%%%%%%%%
%%%%%%%%%%%%%%%%%%%%%%%%%%%%%%%%%%%%%%%%%%%%%%%%%%%%%%%%%%%%%%%%%%

\vspace{10pt}

\begin{figure}[htbp]
\centering
\includegraphics[width=0.7\linewidth]{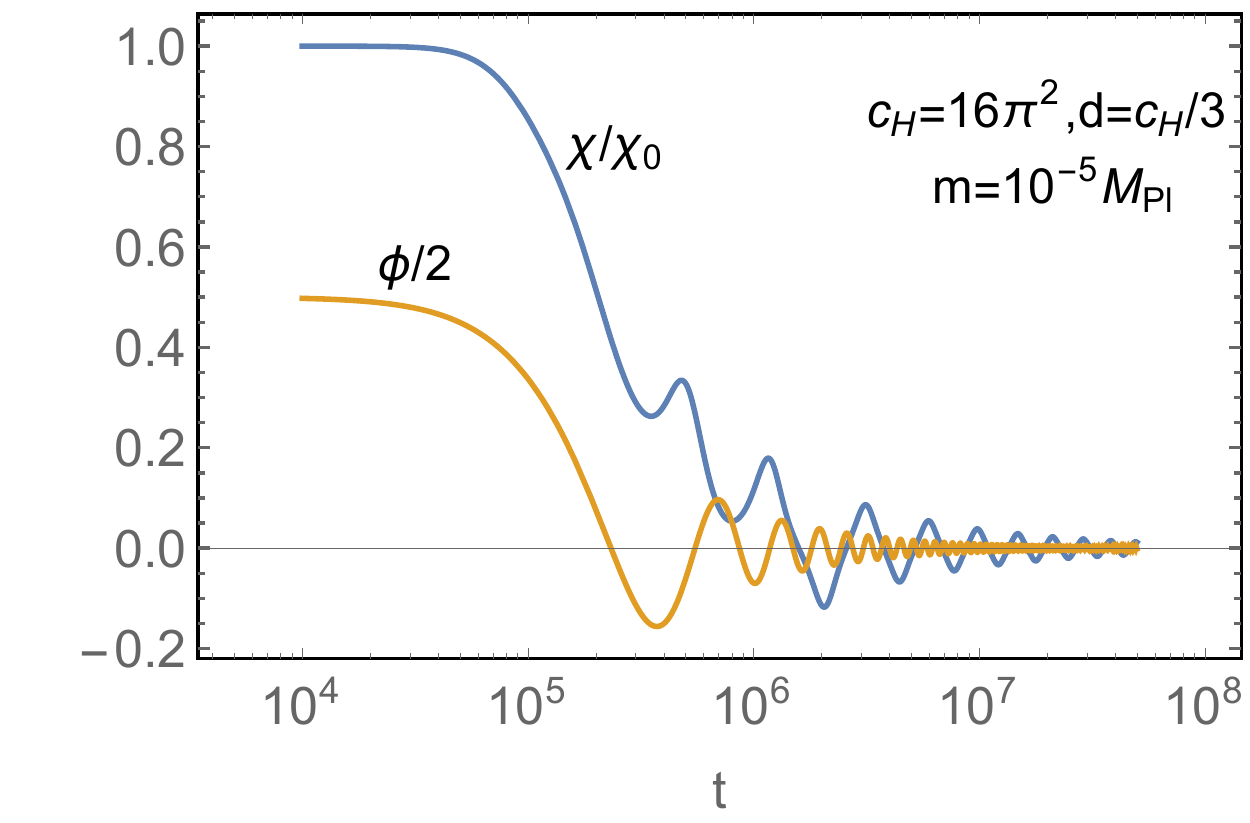}
\caption{\sl \small
The similar figure as Figure~\ref{fig:osc_4} but with a coupling of the PQ  breaking field to the kinetic term of the inflaton.
}
\label{fig:osc_4_match}
\end{figure}

Let us discuss whether the beginning of the oscillation around the origin can be delayed so that $P_\text{osc} \ll P_\text{inf}$.
The reason why the PQ breaking field starts oscillation is that $\rho_i \neq \rho_0$ [see Sec.~\ref{sec:analytic}]
due to the rapid change of the Hubble induced mass term of the PQ breaking field after inflation.
%If a coupling between the inflaton and the PQ breaking field forces $\rho_i \simeq \rho_0$,
%the beginning of the oscillation is delayed.
%For this purpose,
To suppress the oscillation,
we introduce a coupling between the PQ breaking field and the kinetic term of the inflaton,
\begin{align}
{\cal L } =  \frac{d}{2} \partial^\mu\phi \partial_\mu \phi \times |P|^2.
\end{align}
If $d\simeq c_H/3$, the Hubble induced mass term varies after inflation slowly
% changes slowly
%after inflation
%does not change drastically after inflation ends
and hence we may obtain $\rho_i \simeq \rho_0$.
We further assume that the Hubble induced mass is large, which may help the PQ breaking field to follow the point $\rho_0$
%potential minimum
adiabatically~\cite{Linde:1996cx,Nakayama:2011wqa}. 
In Figure~\ref{fig:osc_4_match}, we show the evolution of the PQ breaking field for $m=10^{-5}\mpl$, $c_H=16\pi^2$ and $d=c_H/3$.
Still, the PQ breaking field starts oscillation around the origin when $\chi\sim 10^{-1} \chi_0$.
Thus, the oscillation is inevitable for a quartic potential.%
\footnote{
We find that a larger Hubble induced mass allows for smaller $P_\text{osc}$.
However, in that case, the PQ breaking field value during inflation is required to be small so that the PQ breaking field does not dominate the potential energy during inflation.}
%%

%%%%%%%%%%%%%%%%%%%%%%%%%%%  Figure 6 %%%%%%%%%%%%%%%%%%%%%%%%%%%
%%%%%%%%%%%%%%%%%%%%%%%%%%%%%%%%%%%%%%%%%%%%%%%%%%%%%%%%%%%%%%%%%%

%%%%%%%%%%%%%%%%%%%%%%%%%%%%%%%%%%%%%%%%%%%%%%%%%%%%%%%%%%%%%%%%%%%

\subsection{Sextet and quartic potential}

As we have discussed in Sec.~\ref{sec:analytic}, the growth of fluctuations  is avoided if the oscillation of the PQ breaking field after inflation is driven by a sextet potential.
We assume the following potential of the PQ breaking field $P$,
\begin{align}
\label{eq:PQ potential 46}
V(P) = -m_P^2 |P|^2 + \lambda_4^2 |P|^4  + \lambda_6^2  |P|^6,
\end{align}
where $m_P^2$ and, $\lambda_4$ and $\lambda_6$ are constants.
We again assume the interaction with the inflaton in Eq.~(\ref{eq:potential}).
Then the PQ breaking field obtains a large field value during inflation.
We assume that $\lambda_6$ is sufficiently large so that the field value of the PQ breaking field during inflation is determined by a balance between the sextet term and the Hubble induced mass as
\begin{align}
\label{eq:PQ inf 46}
P_\text{inf} =
\frac{c_H^{1/4} H_\text{inf}^{1/2}}{3^{1/4}\lambda_6^{1/2}}  = \mpl c_H^{1/4} \left(\frac{H_\text{inf}}{10^{14}\text{GeV}}\right)^{1/2} \left( \frac{ 2\times 10^{-5} \mpl^{-1}  }{\lambda_6}\right)^{1/2}.
\end{align}

After inflation, the PQ breaking field starts oscillation driven by the sextet term.
During the oscillation by the sextet term, the parametric resonance effect is ineffective and hence the PQ symmetry is not restored.
The oscillation of the PQ breaking field is eventually driven by the quartic term
when it dominates over the sextet term at $P=P_\text{osc-4}$, 
\begin{align}
P_\text{osc-4} = \frac{\lambda_4}{\lambda_6} = 10^{16}\text{ GeV} \times \frac{\lambda_4}{7\times 10^{-8}} \frac{2\times 10^{-5} \mpl^{-1} }{\lambda_6}.
\end{align}
As long as $P_\text{osc-4} < 10^4 \vev{P}$, the PQ symmetry is not restored and hence domain walls are not formed.
Therefore, this model can solve both isocurvature and domain wall problems for high scale inflation models. 

Before closing this subsection, let us discuss the VEV of the PQ breaking field $\vev{P}$.
If $\lambda_4$ is sufficiently small, 
the VEV  is determined by a balance between
the sextet and the quadratic terms.
If not, the VEV is determined by a balance between
the  quartic and the quadratic terms;
\begin{align}
\vev{P} =
\left\{
\begin{array}{ll}
  \frac{m_P^{1/2}}{3^{1/4} \lambda_6^{1/2}} = 1\times 10^{12}\text{GeV} \times \left(\frac{2\times 10^{-5}\mpl^{-1}}{\lambda_6}\right)^{1/2} \left(\frac{m_P}{20 \text{GeV}}\right)^{1/2} 
   &(\lambda_4 < \sqrt{\lambda_6 m_P}),\\[1.5em]
\frac{1}{ \sqrt{2}\lambda_4} m_P = 1\times 10^{12}\text{GeV} \times \frac{7\times 10^{-8}}{\lambda_4} \frac{m_P}{10^{5}\text{GeV}}
  &(\lambda_4 > \sqrt{\lambda_6 m_P}).
\end{array}
%= 9\times 10^{-9}  \times \left( \frac{\lambda_6}{2\times 10^{-5} \mpl^{-1}}\right)^{1/2} \left( \frac{m_P}{10^{5}\text{GeV}}\right)^{1/2} .\nonumber
%\lambda_6 < \frac{\lambda_4^2}{ m_P} =   \mpl^{-1} \left(\frac{\lambda_4}{7\times 10^{-8}}\right)^2 \frac{10^{5}\text{GeV}}{m_P}.\nonumber 
\right.
\end{align}

In the former case, the mass of the PQ breaking field, $m_P$, is given by $P_\text{inf}$, $\vev{P}$ and $H_\text{inf}$ as
\begin{align}
\label{eq:mass}
m_P = c_H^{1/2} H_\text{inf} \frac{\vev{P}^2}{P_\text{inf}^2} =10\text{GeV} \times c_H^{1/2}\frac{H_\text{inf}}{10^{14}\text{GeV}}
\left(\frac{\vev{P}}{10^{12}\text{GeV}}\right)^2 \left(\frac{\mpl}{P_\text{inf}}\right)^2.
\end{align}
In the latter case,
expressing $\lambda_4$ and $\lambda_6$ by $\vev{P}$ and $P_\text{inf}$ respectively, and put the constraints $\lambda_4^2  > \lambda_6 m_P $ and $P_\text{osc-4} < 10^4 \vev{P}$, we obtain the bound on $m_P$,%
\footnote{
If one adds a potential with a larger power than six instead of the sextet potential, the upper bound of $m_P$ becomes severer.
}
\begin{align}
\label{eq:mass range}
10~\text{GeV}\times c_H^{1/2} \left(\frac{\vev{P}}{10^{12}\text{GeV}}\right)^2 \frac{H_\text{inf}}{10^{14}\text{GeV}} \times  \left(\frac{\mpl}{P_\text{inf}}\right)^2<m_P  \nonumber \\
<10^{5}~\text{GeV}\times c_H^{1/2} \left(\frac{\vev{P}}{10^{12}\text{GeV}}\right)^2 \frac{H_\text{inf}}{10^{14}\text{GeV}} \times  \left(\frac{\mpl}{P_\text{inf}}\right)^2.
\end{align}
%%

%%%%%%%%%%%%%%%%%%%%%%%%%%%%%%%%%%%%%%%%%%%%%%%%%%%%%%%%%%%%%%%%%%%

\subsection{Embedding in supersymmetric theory}

Let us discuss the embedding of our model in supersymmetric theory.
We introduce three chiral multiplets $P$, $X$ and $Y$, whose PQ charges are $1$, $-2$ and $-3$, respectively.
Then the superpotential of them is given by 
\begin{align}
W = \lambda_4 XP^2 + \lambda_6 Y P^3,
\end{align}
where $\lambda_4$ and $\lambda_6$ are constants.
The F terms of $X$ and $Y$ yield the quartic and the sextet terms of $P$ in Eq.~(\ref{eq:PQ potential 46}), respectively.
$m_P^2$ is provided by a soft supersymmetry breaking mass term of $P$.
Assuming that the Hubble induced mass terms as well as soft supersymmetry breaking mass terms of $X$ and $Y$ are positive, they are fixed to their origin during and after inflation.
Then, we obtain the potential in Eq.~(\ref{eq:PQ potential 46}).

It should be noted that the allowed range of the mass of the PQ breaking field in Eqs.~(\ref{eq:mass}) and (\ref{eq:mass range}) is consistent with supersymmetry breaking and its mediation by Planck-suppressed interactions.
With the mediation by Planck-suppressed interactions,
the fermion superpartner of axion, axino, also has mass of order $m_P$.
If is known that the axino with mass less than $O(10^4)\text{ GeV}$,
causes cosmological difficulties~\cite{Kawasaki:2007mk}.
However, for $m_P \simeq 10^5$ GeV which is allowed for large $\lambda_4$ 
[Eq.(\ref{eq:mass range})], neither the saxion nor the axino brings about cosmological problems.
%soft scalar masses of supersymmetric particles are also as large as $m_P$.
% In particular, for $m_P \simeq 10^5$ GeV, the saxion nor the axino bring about cosmological problems~\cite{Kawasaki:2007mk}.

Finally, we note that
soft scalar masses of $\sim 10^5$ GeV fit in well with high scale supersymmetry breaking scenarios~\cite{Giudice:1998xp,Wells:2004di,Ibe:2006de,Hall:2011jd,ArkaniHamed:2012gw}.
There, the observed Higgs mass of $125$ GeV~\cite{Aad:2012tfa,Chatrchyan:2012ufa} is explained by large quantum corrections from scalar tops with masses of $10^5$ GeV~\cite{Okada:1990vk,Ellis:1990nz,Haber:1990aw}.
Since anomaly mediation generates gaugino masses as large as $1$~TeV~\cite{Giudice:1998xp,Randall:1998uk} (see also Refs.~\cite{Dine:1992yw}),
the supersymmetry breaking field does not have to be a singlet field.
Then the Polonyi problem~\cite{Coughlan:1983ci,Ibe:2006am} is absent.
For the gravitino mass is as large as $10^5$ GeV,
the gravitino decays before the Big-Bang-Nucleosynthesis begins.
Thus, our model of PQ symmetry breaking, combined with high scale supersymmetry breaking scenarios, is free from cosmological problems.

\section{Summary and discussion}

In this paper, we have discussed the solution to the isocurvature perturbation problem of the PQ mechanism by
a large PQ breaking scale during inflation.
We have investigated the evolution of the PQ breaking field after inflation and discussed whether the PQ symmetry restores after inflation, which leads to the formation of domain walls.
We have proposed a model without the restoration of the PQ symmetry.
%We have found that the restoration does not occur if the quartic potential of the PQ breaking field is absent.
%In that case, however, the mass of the PQ breaking field must be as large as $O(10)$ GeV, which causes cosmological moduli problems.
%We have proposed a model with a sufficiently large mass of the PQ breaking field.
Interestingly, the predicted mass of the PQ breaking field is not very far from the electroweak scale.
Thus, our model is compatible with a supersymmetric theory where the mass of the PQ breaking field is given by supersymmetry breaking.

As we have discussed in Sec.~\ref{sec:review},
the field value of the PQ breaking field must be as large as the Planck scale,
if the Hubble scale during inflation is of $O(10^{13})$ GeV.
It is known that a field which obtains a Plank scale field value by a Hubble induced mass affects the prediction of large field inflation models~\cite{Harigaya:2015pea}.
With this effect, the chaotic inflation model simply driven by a mass term of the inflaton is consistent with recent observations of the CMB~\cite{Ade:2015lrj}.

\section*{Acknowledgements}
This work is supported by Grants-in-Aid for Scientific Research from the Ministry of Education, Culture, Sports, Science, and Technology (MEXT), Japan,
No. 24740151 and No. 25105011 (M.~I.),
No. 25400248 (M.~K.)
as well as No. 26104009 (T.~T.~Y.); 
Grant-in-Aid No. 26287039 (M.~I. and T.~T.~Y.) from the Japan Society for the Promotion of Science (JSPS); and by the World Premier International Research Center Initiative (WPI), MEXT, Japan.
K.H. is supported in part by a JSPS Research Fellowship for Young Scientists.

\end{document}